\documentstyle[prd,aps,twocolumn,epsfig]{revtex}
\def\br{\begin{eqnarray}}
\def\er{\end{eqnarray}}
\def\be{\begin{equation}}
\def\ee{\end{equation}}

\def\g{\gamma}

\def\m{\mu}

\def\({\left(}
\def\){\right)}
\def\<{\left\langle}
\def\>{\right\rangle}

\def\S{\Sigma}

\begin{document}
\twocolumn[\hsize\textwidth\columnwidth\hsize\csname %%% TWO COLUMN
@twocolumnfalse\endcsname                            %%% TWO COLUMN
%
%\draft
%
\title{Limit on the fermion masses in technicolor models }
\author{
A. Doff and A. A. Natale\\
}
\address{
Instituto de F\'{\i}sica Te\'orica, UNESP, Rua Pamplona 145, 01405-900, S\~ao Paulo, SP, Brazil}
%%%%
\date{\today}
\maketitle
%%%%
\begin{abstract}
\par Recently it has been pointed out that no limits can be put on the scale of fermion mass generation
$(M)$ in technicolor models, because the relation between the fermion masses $(m_f)$ and $M$ depends
on the dimensionality of the interaction responsible for generating the fermion mass. Depending on this
dimensionality it may happens that $m_f$ does not depend on $M$ at all. We show that exactly in this
case $m_f$ may reach its largest value, which is almost saturated by the top quark mass. We make
few comments on the question of how large can be a dynamically generated fermion mass.
\end{abstract}
 \vskip 0.2cm]
%\section{Introduction}
\par The mechanism that breaks the electroweak gauge symmetry, $SU(2)_L \times U(1)_Y$, down to the gauge
symmetry of electromagnetism $U(1)_{em}$ is the only still obscure part of the standard model. It is known that
up to the scale of $1$ TeV some sign of this mechanism has to become manifest in future experiments.
In the same way that an upper bound on the scale of the electroweak symmetry breaking has been put
forward (the  $1$ TeV scale), it was thougth that the scale of fermion mass generation had also an upper bound,
and that bound would be at reach of the next generation of accelerators \cite{ac}.
Recently it was shown that no upper bound can be put on the scale of fermion mass generation beyond that on
the scale of electroweak symmetry breaking \cite{mnw}. This result was obtained considering the scattering of
same helicity fermions into a large number of longitudinal weak vector bosons in the final state, and it was also
obtained in a more involved way in Ref.\cite{clt}.  This result is important and at the same time is disappointing,
because an upper bound on the scale of fermion mass generation would provide a target for future accelerators
in order to understand the origin of fermion masses.
The scale related to the origin of fermion masses cannot  be bounded but the fermion mass itself is bounded. The
bound on the fermion masses comes out from the upper limit on the Yukawa coupling $(\lambda_y \leq \sqrt{8\pi})$
\cite{yuk}.  In the standard scenario this is not very interesting because it also leads to a bound in the fermion
masses of the order of $1$ TeV. Therefore there is still space for a heavy new family (respecting the constraints
provided by the high precision experiments). This problem becomes much more interesting in theories with
dynamical symmetry breaking like technicolor theories,  where, in principle, some of the free parameters of the
standard model are calculable as long as we know the symmetries of the underlying theory that is responsible
for the mass generation.
Let us recall some of the arguments about the nonexistence of a bound on the scale for fermion mass generation
in technicolor models \cite{mnw}. In these models the fermion mass is given by
\be m_f \approx c \frac{\< \overline{\psi}_{tc}\psi_{tc} \>}{M^2_{etc}}  , \label{mf} \ee
where $c$ is a constant and $ \<\overline{\psi}_{tc}\psi_{tc} \>$ is the technifermion condensate. $M _{etc}$ is
the mass of the extended technicolor boson and is the mass scale that reproduces an effective Yukawa coupling.
According to Ref.\cite{ac}  $M _{etc}$    should be bounded in the following way. If technicolor is a QCD-like
model we can assume   $ \< \overline{\psi}_{tc}\psi_{tc} \> \approx v^3$ \cite{hg} ,  where $v \approx 246$ GeV,
and assuming $c \simeq g^2_{etc}$ we obtain
\be M _{etc}   \approx \left( g^2_{etc} \frac{v^3}{m_f} \right)^{1/2}\hspace{-0.5cm}, \label{metc} \ee
which gives the bound on the mass scale responsible for fermion mass generation. Nowadays it is known that
composite operators like $\< \overline{\psi}_{tc}\psi_{tc} \>$ (and the technifermion self energy) may have a
large anomalous dimension ($\gamma_m \gtrsim 1$) in such a way that the fermion mass is given by
\be m_f \approx c \frac{\< \overline{\psi}_{tc}\psi_{tc} \>}{M^2_{etc}} \left(\frac{M^2_{etc}}{v^2}
\right)^{\gamma_m } , \label{mfg} \ee
from where we notice that for $\gamma_m =1$ there is no relation at all between $m_f$ and $M _{etc}$, indicating
the inexistence of a bound on this last mass scale.
The anomalous dimension $\gamma_m =1$ can be
obtained in the extreme limit of a walking technicolor dynamics \cite{holdom}, corresponding to a near critical extended
technicolor interaction with increased importance of four-fermion operators, and, if the $M_{etc}$ scale is
raised,  $\gamma_m =1$ possibly only happens with a fine tuning of the theory.
Let us still continue to discuss the case  $\gamma_m =1$.
Exactly in this case we cannot establish a bound on  $M _{etc}$, but note that it also imply that the maximum
dynamical fermion mass is limited by
\be m_f  \leq c v . \label{mfgmax} \ee
 In this note we propose to discuss which is the maximum value admited by
Eq.(\ref{mfgmax}) or by the dynamical fermion mass in general. We will compute the dynamical fermion mass
described in Fig.(1), where the ordinary fermions ($f$) are connected to technifermions ($T_f$) through an
extended technicolor gauge boson associated to some gauge group ($SU(N_{etc})$ with coupling
$\alpha_{etc}=g^2_{etc}/4\pi$.)
\vskip -0.1cm
\begin{figure}[htb]
\begin{center}
\epsfig{file=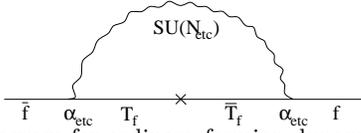,width=0.27\textwidth}\label{diagmassa} \caption{Diagram for ordinary fermion
dynamical masses in technicolor models.}
\end{center}
\end{figure}
We perform the calculation of Fig.(1) using the following general expression for the techniquark self-energy
\cite{dn}
 \be \S(p)_{g} = \mu\(\frac{\mu^2}{p^2}\)^{\theta}[1 +
bg^2_{tc}(\mu^2)ln(p^2/\mu^2)]{,}^{\!\!\!-\gamma{\cos(\theta\pi)}} \label{mger1} \ee
\noindent where in the last equation we identified $\gamma = \gamma_{tc}$.
The scale $\m$ (or $v$) is related to the technicolor condensate  $\< \bar{\psi_{tc}}\psi_{tc} \> \equiv \mu^3$
and is ultimately fixed by the experimental value of the weak gauge boson masses.
The advantage of using such
expression is that it interpolates between the extreme possibilities for the technifermion self-energy, {\it i.
e.} when $\theta=1$ we have the soft self-energy giving by
\be \hspace{0.5cm}\S_s (p) = \frac{\mu^3}{p^2}[1 + bg_{tc}^2(\mu^2)ln(p^2/\mu^2)]^{\g}, \label{eq2} \ee which is
the one obtained when the composite operator  $\< \overline{\psi}_{tc}\psi_{tc} \> $ has canonical
dimension. When $\theta=0$ operators of higher dimension may lead to the hard self-energy expression
\be \S_h (p) = \mu[1 + bg^2_{tc}(\mu^2)ln(p^2/\mu^2)]^{-\g}, \label{einc} \ee where $\g$ must be larger than
$1/2$ and the self-energy behaves like a bare mass \cite{kl}. Therefore no matter is the dimensionality of the
operators responsible for the mass generation in technicolor theories the self-energy can always be described by
Eq.(\ref{mger1}). In the above equations $g_{tc}$ is the technicolor coupling constant and $\g
=3C_{tc}/16\pi^2b$, where $C_{tc}=\frac{1}{2}[C_2(R_{1}) + C_2(R_{2}) - C_2(R_{\overline{\psi}\psi})]$ , with
the quadratic Casimir operators $C_2(R_{1})$ and $C_2(R_{2})$ associated to the $R.H$ and $L.H$ fermionic
representations of the technicolor group, and $C_2(R_{\overline{\psi}\psi})$ is related to the condensate
representation. $b_{tc}=\frac{1}{16\pi^2}[11N - \frac{2}{3}n_{f}]$  is the $g_{tc}^3$ coefficient of the
technicolor group $\beta$ function. The complete equation for the dynamical fermion mass is \be m_{f} =
\,\frac{3C_{etc}\mu}{16\pi^4}\int\,dq^4\left(\frac{\mu^2}{q^2}\right)^{\theta}\frac{g^{2}_{etc}(q)[1 +
b_{tc}g^2_{tc}ln(\frac{q^2}{\mu^2})]^{-\delta}}{(q^2 + M^{2}_{etc})(q^2 + \mu^2)},\label{eqmf} \ee where
$C_{etc}$ is the Casimir operator related to the etc fermionic representations, a factor $\mu$ remained in the
fermion propagator as a natural infrared regulator and $\delta =\gamma{\cos\theta\pi}$, $g^{2}_{etc}(q)$ is
assumed to be giving by
\be g^2_{etc}(q^2) \simeq \frac{g_{etc}^2(M^2_{etc})}{(1 +
b_{etc}g^2_{etc}(M^2_{etc})ln(\frac{q^2}{M^2_{etc}}))} . \label{eqgetc} \ee
Note that in Eq.(\ref{eqmf})  we have two  terms of the form $[1+b_i g^2_i ln q^2]$ where the index $i$ can be
related to $tc$ or $etc$. To obtain an analytical formula for the fermion mass we will  consider  the
substituition $q^2 \rightarrow \frac{xM^2_{etc}}{\m^2}$, and we will  assume that $b_{etc}g^2_{etc}(M_{etc})
\approx b_{tc}g^2_{tc}(M_{etc})$ , what will simplify considerably the calculation. Knowing that the $etc$ group
usually is larger than the $tc$ one we computed numerically the error in this approximation for few examples
found in the literature. The resulting expression for $m_f$ will be overestimated by a factor $1.1 - 1.3$ and is
giving by
 \be m_{f} \simeq\ \,\frac{3C_{etc}g^{2}_{etc}(M_{etc})\mu}{16\pi^2}\left(\frac{\mu^2}{M^2_{etc}}\right)^{\theta}\left[1 +
b_{tc}g^2_{tc}ln\frac{M^2_{etc}}{\m^2}\right]^{-\delta}\!\!\!\!\!\!I, \label{eqfin}
 \ee
 where
$$
I = \frac{1}{\Gamma(\sigma)}\int_{0}^{\infty}\!\!\!d{\sigma}\sigma^{\epsilon -1}e^{-\sigma}\frac{1}{\theta +
\alpha\sigma},
$$
and  $\epsilon = \delta + 1 = \gamma {cos\theta\pi} + 1$, $\alpha=b_{tc}g^2_{tc}(M_{etc})$.
 To obtain Eq.(\ref{eqfin}) we made use of the following Mellin transform
 \be \left[ 1 + \kappa \ln \frac{x}{\mu^2} \right]^{-\epsilon} \!\!\!\!=
\frac{1}{\Gamma(\epsilon)}\int_0^\infty d\sigma \, e^{-\sigma} \left( \frac{x}{\mu^2} \right)^{-\sigma
\kappa}\!\!\!\!\!\! \sigma^{\epsilon - 1}. \label{eqx} \ee
\noindent Finally we obtain
 \be m_{f} \simeq\ \,\frac{3C_{etc}g^{2}_{etc}(M_{etc})\mu}{16\pi^2}\left(\frac{\mu^2}{M^2_{etc}}\right)^{\theta}F(\cos\theta\pi,\gamma,\alpha).\label{eqfin2}\ee
\noindent where
\br F(\cos\theta\pi,\gamma,\alpha) = &&\left[1 + b_{tc}g^2_{tc}ln\frac{M^2_{etc}}{\m^2}\right]^{-\gamma
{\cos(\theta\pi)}}\hspace{-1.2cm}\Gamma(-\gamma
\cos(\theta\pi),\theta/\alpha)\nonumber\\
&&\exp({\frac{\theta}{\alpha}})\alpha^{-1-\gamma \cos(\theta\pi)}\theta^{\gamma\cos(\theta\pi)}. \nonumber\er
 \par Simple inspection of the above equation shows that (as long as $ M_{etc} > \mu$) the largest value for the
 fermion mass happens for  $\theta = 0$, and expanding Eq.(\ref{eqfin2}) near this point we have
\br m_{f} \simeq\ &&\,\frac{3C_{etc}g^{2}_{etc}(M_{etc})\mu}{16\pi^2}\left[1 +
b_{tc}g^2_{tc}ln\frac{M^2_{etc}}{\m^2}\right]^{-\gamma}\times\nonumber\\
&&\frac{1}{\gamma {b_{tc}g^2_{tc}(M_{etc})}} (1 +  O (\theta) + ... ) \label{eq23}
 \er
 and for   $\theta = 0$ we obtain
 \br m_{f} \simeq\ \,\frac{C_{etc}g^{2}_{etc}(M_{etc})\mu}{C_{tc}g^2_{tc}(\mu)}\left[1 +
b_{tc}g^2_{tc}ln\frac{M^2_{etc}}{\m^2}\right]^{-\gamma + 1} \label{eqthe0}
 \er
 which gives the largest dynamical fermion mass that we can generate. Although this result is simple and quite
 intuitive we have not been able to find it stated anywhere.
\par Note that using the expression for the running coupling $g^2_{etc}(M_{etc})$ Eq.(\ref{eqthe0}) can be written in the following form:
\be m_{f}\sim \frac{C_{etc}}{C_{tc}}\left(\frac{\alpha_{etc}}{\alpha_{tc}}\right)^{\gamma_{tc}}\!\!\!\!\!\!\mu
\sim cv \label{eq16}\ee \noindent where $\gamma$ of the previous expressions is indicating $\gamma_{tc}$, and
the factor $c$ is now giving by $c
=\frac{C_{etc}}{C_{tc}}\left(\frac{\alpha_{etc}}{\alpha_{tc}}\right)^{\gamma_{tc}}$. The possible values of $c$ will
determine the maximum value of the fermion mass.
 To find some limits on the dynamical fermion mass let us consider some possible ways to introduce the
 extended technicolor theory. We may, for example, consider that the $etc$ theory may be a kind of
 grand unified theory (gut) based on the group $SU(k)$ containing technicolor and the standard Georgi-Glashow
 group \cite{hgg}, we than have
$$
SU(k) \supset SU(k-5)_{tc}\otimes SU(5)_{gg} ,
$$
\noindent  where $SU(k-5)_{tc}$ is the $tc$ group, and $SU(5)_{gg}$ is the gut of Ref. \cite{hgg}.
As $tc$ is a strongly interacting theory it is natural to have  $k \geq 7$. Therefore, associating the
$SU(k)$ group to etc, we obtain the following ratio of Casimir operators
$$
\frac{C_{etc}}{C_{tc}} = \frac{(k^2 -1 )(k-5)}{k((k-5)^2-1)} .
$$
\noindent On the other hand we must also preserve asymptotic freedom, what imply $k \leq 11$
\cite{hg}\cite{ref4},  and the ratio $r_{c}=\frac{C_{etc}}{C_{tc}}$ will take values in the range $ r_{c} =  1.7
- 4.5$. We still have to look at the ratio of coupling constants. As $tc$ is a QCD like theory we can assume as
usual that $\alpha_{tc} \sim 1$. The $etc$ theory can be associated to a gut in this case. Actually, there is no
reason at all (specially when the self-energy is the expression with $\theta=0$) to expect a low value for
$M_{etc}$ and a natural one could be $M_{etc}=\Lambda_{gut} \sim 10^{16}$ Gev with $\alpha_{etc} \approx \alpha_{gut}
\sim (40)^{-1}$. The coefficient $\gamma_{tc}=\frac{3C_{tc}}{16\pi^2b_{tc}}$  must be larger than $0.5$. and in
fact  if the $tc$ group is $SU(2)_{tc}$ we have $\gamma_{tc}\sim 0.5$, for other (and larger) models this
coefficient will be larger than $1/2$. Therefore we roughly have
$$
\left(\frac{\alpha_{etc}}{\alpha_{tc}}\right)^{\gamma_{tc}} \sim \left(\frac{1}{40}\right)^{\frac{1}{2}} \sim
\frac{1}{6}.
$$
Finally, considering all the estimates we obtain
\be m_{f}^{max} \sim O(0.3 - 0.8)v \sim O(75 - 200) Gev.
 \label{mlim}
 \ee
\par Note that this is a rough estimative and possibly is the best that we can do considering the
present knowledge of strongly interacting theories. Our calculation is possibly overestimated and it
should be divided by a factor $1.1$ to $1.3$ as we informed in the paragraph after Eq. (\ref{eqgetc}).
We also assumed an extreme case for the self-energy maximazing the fermion mass and it is not clear
if a realistic model can reproduce exactly this behavior. Therefore  considering only the smaller factor
($1.1$) discussed above it seems that the maximum value of the dynamical mass is already saturated by the
top quark mass.
There is a possible way to circumvent this limit, {\it i.e.} we could build a model with  a fermion
more massive than the limit given by Eq.(\ref{mlim}) where the mass comes from the contribution
of several diagrams. In this case the fermion mass could be given by $m_f = n m_f^{max}$, where
$n$ is the number of diagrams contributing to the mass of one specific fermion. Models of this kind
are similar to the ones of Ref.\cite{ref2} (a  $SU(9)_{gut}\otimes SU(3)_H$ theory, with a technicolor gut and a
horizontal symmetry group), which is based on the model of
Ref.\cite{ref1} (a $SU(7)$ technicolor gut).  In the table below we show the maximum fermion
mass that we can obtain in such models. In the  $SU(9)$ model we have two diagrams feeding up
the heaviest fermion, even so it is difficult to obtain a mass larger than the limit of Eq.(\ref{mlim}).
Note also that this result is quite dependent on the model and the introduction of a horizontal symmetry
is necessary for building a realistic model and to give several contributions to the fermion masses.
\begin{center}
\begin{tabular}{|c|c|c|c|c|}
  \hline
  $SU(k)$ & $r_{c}={C_{etc=gut}}/{C_{tc}}$ & $\gamma_{tc}$ &  $n$  & $m_{f}^{max}$ \\
  \hline
  $SU(7)$ & 4.5 & 0.50 &  1 & $O(177)$ $Gev$ \\
  $SU(9) $& 2.4 & 0.65 &  2 & $O(110)$ $Gev$\\
  \hline
\end{tabular}
\end{center}
\par We can also consider a different class of models where the $etc$ group  ($G_{etc}$) and the
standard model ($G_{sm} $)  obey the following \cite{ref5}
\be G_{etc}\otimes G_{sm} =  SU(N_{etc})\otimes SU(3)_{c}\otimes SU(2)_{L}\otimes U(1)_{y}, \ee
\noindent where  $SU(N_{etc})$ must be large enough to accommodate technicolor.   No realistic model
has yet been found along this line, but let us consider  a model based on the
$SU(5)_{etc}$ group \cite{ref6}\cite{ref7}, which contain $SU(2)_{tc}$ and one technifermion generation.
To obtain the hierarchy among the 3 generations the model has the following symmetry breaking
structure
\br &&SU(5)_{etc} \otimes G_{sm}\nonumber \\&&\,\,\,\,\,\downarrow\,\,\Lambda_{1}\sim 1000\,Tev\nonumber\\
&&SU(4)_{etc}\otimes G_{sm}\nonumber\\&&\,\,\,\,\,\downarrow\,\,\Lambda_{2}\sim 100\,Tev\nonumber\\&&SU(3)_{etc}\otimes G_{sm}\nonumber \\
&&\,\,\,\,\,\downarrow\,\,\Lambda_{3}\sim 10\,Tev\nonumber\\&&SU(2)_{tc} \otimes G_{sm} \er
\par  We will not discuss the details of this model but just assume that the $tc$ dynamics has
the behavior of Eq.(\ref{eqthe0}) and compute the mass of the heaviest family wich is given by
\be  m_{3} \sim
\frac{C_{3}}{C_{tc}}\left(\frac{\alpha_{3}(\Lambda_{3})}{\alpha_{tc}}\right)^{\gamma_{tc}}\hspace{-0.5cm}v \sim c_{3}\,v.\nonumber \\
\ee With the values discussed in Ref.\cite{ref6}\cite{ref7} we see that we do not obtain a very large mass and
the limit of Eq.(\ref{mlim}) seems to be common to all models.
It is also interesting to discuss another kind of constraint that can be put on the dynamical fermion masses.
Fermion masses have been
limited in the standard model making the analysis of the partial wave amplitudes ($J=0$) of the processes
${\bar{f}} f \rightarrow  {\bar{f}} f$ at high energies. These amplitudes will be proportional to $a_0 \propto
m_f^2 G_F$, and the unitarity condition of the S matrix imply that $|a_0| \leq 1$, which gives the bound
\cite{ref8}
\be m_f^2 \lesssim \frac{2\pi \sqrt{2}}{3G_F}. \label{limchi} \ee
The question that we address now is if this limit can be directly applied to $tc$ theories. If we follow
Ref.\cite{csoni} it seems that this result could be applied also to $tc$ theories. In Ref.\cite{csoni} it was
shown that one technicolor theory where the dynamical symmetry breaking is generated due to the effect of higher
order operators the resulting effective theory reproduces exactly the standard model (their self-energy solution
is identical to the one we are discussing here). The gauge interactions of the ordinary fermions with the
standard gauge boson is obviously the same, but more importantly the Higgs boson coupling is also reduced to the
standard model one, what is fundamental to obtain the result of Eq.(\ref{limchi}). Only a  light degree of
freedom (a scalar composite boson) appears bellow the TeV scale. Therefore the limit of Eq. (\ref{limchi}) could
be valid for technicolor when the $tc$ dynamics is the harder one that is discussed here. Of course, this is not
true for a softer self-energy solution, when the fermion mass is smaller due to the dependence on the scale
$M_{etc}$. If we impose the limit of Eq.(\ref{limchi}) over Eq.(\ref{eq16}) we obtain
\be \frac{\alpha_{etc}(M_{etc})}{\alpha_{tc}(\m)} \lesssim \left(\frac{2\pi
\sqrt{2}}{3G_F}\frac{1}{v^2r^2_{c}}\right)^{\frac{1}{2\gamma_{tc}}}\!\!\!\!. \ee
Considering the  numeric  values $G_{F}\sim1.166\times 10^{-5}GeV^{-2}\!\!,$ $v \sim 246\,Gev$ and the ratio
$r_{c}= 1.7 - 4.5$ we obtain $\alpha_{etc}(M_{etc})/\alpha_{tc}(\m) < 1$. This limit will not imply in a very
strong constraint in the dynamical fermion mass. However there is a problem in the above argument. If the
dynamical symmetry breaking is generated by the effect of higher order operators the effective low energy theory
may reproduce exactly the standard model as claimed in  Ref.\cite{csoni}, but the effective Yukawa coupling will
be also proportional to a form factor which, just on dimensional grounds, should be of the form  $F(q^2
\rightarrow 0) \propto (1 - q^2/\m^2 + ...)$, because any other mass scale (like $M_{etc}$) is erased from the
self-energy (or appears only in a logarithm). Therefore, at low energies the Yukawa coupling is equal to the one
of the standard model, but for momenta $q^2$ near the $tc$ scale ($\m$) this coupling should be quite suppressed
leading to a dynamical fermion mass not higher than $\m$.
\par In conclusion, it seems very difficult to generate dynamical
fermion masses in technicolor models larger than the technicolor scale.
The largest mass that can be obtained appears when we consider the hardest (concerning the momentum dependence)
expression for the technicolor dynamics, which is the same that is also consistent with the nonexistence of a
bound on the scale of fermion mass generation. Maybe models with some extra symmetry (possibly a horizontal
symmetry), implying that the heaviest fermion receives mass contribution from several diagrams, could be one
possibility to have fermions heavier than the top quark within the technicolor scheme, although we do not know
any realistic model along this line. Otherwise, if technicolor is responsible for the standard model symmetry breaking,
 it seems that no other ordinary heavier fermion family will be found in the next generation of accelerators.
\section*{Acknowledgments}
This research was supported by the Conselho Nacional de Desenvolvimento Cient\'{\i}fico e Tecnol\'ogico (CNPq)
(AAN) and by Fundac\~ao de Amparo \`a Pesquisa do Estado de S\~ao Paulo (FAPESP) (AD).

\end{document}